\documentclass[aps,prl,twocolumn,showpacs,superscriptaddress]{revtex4}  
\usepackage{graphicx,color}  
\usepackage{dcolumn}   
\usepackage{bm}        
\usepackage{amsmath,amssymb}   
\usepackage[caption=false]{subfig}

\begin{document}

\widetext


\title{Deterministic extinction by mixing in cyclically competing species}
\author{Cilie W. Feldager}
\affiliation{
  Center for Models of Life, Niels Bohr Institute, University of Copenhagen, Blegdamsvej 17, 2100 Copenhagen, Denmark.}
\author{Namiko Mitarai}
\email{mitarai@nbi.dk}
\affiliation{
  Center for Models of Life, Niels Bohr Institute, University of Copenhagen, Blegdamsvej 17, 2100 Copenhagen, Denmark.}
\author{Hiroki Ohta}
\email{ohta.hiroki.6c@kyoto-u.ac.jp}
\affiliation{
 Niels Bohr International Academy / Center for Models of Life, Niels Bohr Institute, University of Copenhagen, Blegdamsvej 17, 2100 Copenhagen, Denmark.}

\date{\today}

\begin{abstract}
We consider a cyclically competing species model on a ring with global mixing at finite rate, 
which corresponds to the well-known Lotka-Volterra equation in the {limit of} infinite mixing rate.
Within a perturbation analysis of the model from the infinite mixing rate, 
we provide analytical evidence that extinction occurs deterministically
at sufficiently large but finite values of the mixing rate for any species number ${N\ge3}$.
Further, {by focusing on the cases of rather small species numbers,}
we discuss numerical results concerning the trajectories toward such deterministic extinction,
including {global} bifurcations {caused by changing the mixing rate}.
\end{abstract}

\maketitle


\section{Introduction}
One of the central interests in the field of theoretical ecology and game theory
is to understand the mechanism of coexistence and extinction in interacting agents \cite{hofbauer1998theory,szabo2007evolutionary,szolnoki2014cyclic}.
Among many kinds of models for interacting agents,
$N$-species cyclic dominance has been studied intensely
due to rich dynamics and possible long-term coexistence. 
Specifically, {simple models for such cyclic dominance have been introduced based on differential equations \cite{itoh1971boltzmann,may1975nonlinear}
or lattice models \cite{tainaka1988lattice,bramson1989flux,durrett1998spatial},}
{and} could provide a fundamental viewpoint to understand more complicated cases observed in nature
\cite{czaran2002chemical,kerr2002,mathiesen2011ecosystems,frentz2015strongly}. 

\begin{figure}
 \includegraphics[width=0.5\textwidth]{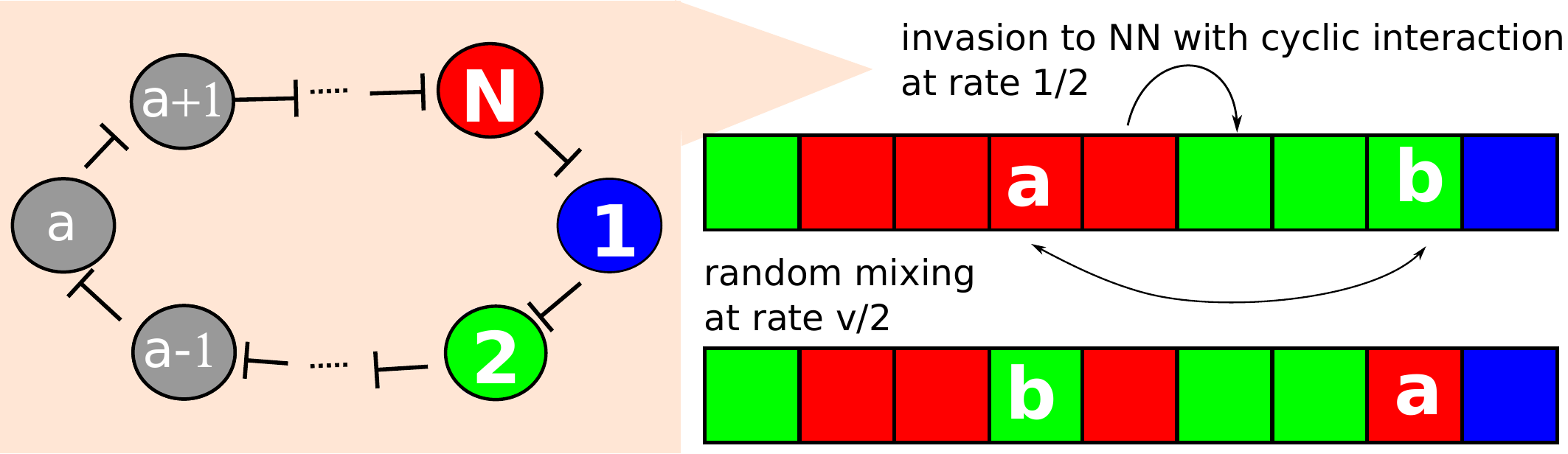}
  \caption{
    \label{ModelFig} (color online)
    A schematic description of the model.
    An invasion through the nearest neighbor (NN) occurs along the cyclic competition.
    Individual species on randomly chosen two sites are exchanged at rate $v/2$.
  }
\end{figure}

Dynamical behaviors of cyclic dominance between interacting agents may be described by the Lotka-Volterra (LV) equation
or a cyclically competing species model on a lattice,
where the former is supposed to be derived in the large size limit of the latter {in} a well-mixed condition.
Specifically, the time evolution of the density $P_{\alpha}$ of a species $\alpha$ for $\alpha\in\{1,2,\cdots,N\}$
in the LV equation is described as follows \cite{hofbauer1998theory}:
\begin{equation}
 \dot{P}_{\alpha} = P_{\alpha}(P_{\alpha_+} - P_{\alpha_-}),
\label{LV}
\end{equation}
where $\dot P_{\alpha}$ is the time derivative of $P_\alpha$. $\alpha_+(\alpha_-)$
denotes a species forward (backward) next to $\alpha$ in the cyclic sense, i.e., it is a prey (predator) of
species $\alpha$ (Fig.~\ref{ModelFig} left panel). 

{One of the} {characteristic properties in the LV equation} {is} {the coexistence of all the species
  where each density shows neutrally stable oscillations with an amplitude given
  by the initial condition \cite{hofbauer1998theory}. }
{The other characteristic properties in LV equation often discussed are qualitative differences between the odd and even species number cases
  \cite{sato2002parity,szabo2007evolutionary}. For example, in the even species number case, odd-labeled species and even-labeled species antagonize each other,
  while in the odd species number case, the overall interaction forms a negative feedback loop}.
{There has been still recent progress on analytical topics in the LV equation, focusing on conserved quantities \cite{Itoh1987,Oleg2009} or the related Lyapunov functions in the case of more general heterogeneous interactions \cite{Zia2011,Frey2013,Haerter2016}.}

On the other hand, a cyclically competing species model on a lattice,
where each species on a lattice stochastically invades
its {neighbor} species along the rule of the cyclic dominance,
shows different behaviors from those in the LV equation even qualitatively.
For example, in the one-dimensional case {with the infinite system size}, it has been proven that when $N\le 4$, 
species at any {site is} invaded after any time {period}, showing slow coarsening, 
while when $N\ge 5$, each site reaches a final state, leading 
{to an absorbing state} \cite{bramson1989flux,fisch1992clustering,frachebourg1996spatial,frachebourg1996segregation}.
Although it has been poorly known how the behaviors of models depend on the dimensionality of the lattices,
an approximation analysis rather suggests
that the behaviors of the models do not approach those in the LV equation
even if the spatial dimension is higher \cite{tainaka1994intrinsic,tainaka1994vortices,frachebourg1998fixation}.

As mentioned above, mixing in lattice models is supposed to bridge the gap between the two descriptions of cyclic ecosystems.
However, if the system is not completely well-mixed,
it could generally produce some macroscopic behaviors qualitatively different from those in the well-mixed condition.
For example, one might consider mixing by local mobility, i.e.,
by locally exchanging the position of two species
{or mixing} induced by a fluid flow,
resulting in spatiotemporal density oscillations such as spirals \cite{reichenbach2007, reichenbach2007mobility}
or density oscillations with an increasing amplitude leading to extinction \cite{karolyi2005rock},
respectively. Thus, such numerical works have already shown that mixing tends to induce oscillatory behaviors.
However, how {a mixed cyclic ecosystem approaches} the LV equation is still not entirely clear. 
In this paper, we propose a simple model where one can extract analytical results, to some extent, on this topic.

{In this paper}, we study a cyclically competing species model on a ring by taking into account
the effects of a global mixing at finite rate.
By performing a perturbative analysis from the large mixing rate limit,
we will show that, at least, one species has to go {\it deterministically} extinct
for {any species number $N\ge 3$ in the large size limit}.
{We call this extinction occurring in the large size limit as deterministic extinction. The deterministic extinction}
is completely different from the extinction {caused by} {\it stochasticity}, {where internal fluctuations as a result of a finite system size
could {cause the extinction of} a species, in particular, when the species population size is low} \cite{reichenbach2006noise}.
It turns out that such deterministic extinction {is} key to understand why the relationship between
the lattice models and the LV-equation for cyclic ecosystems was elusive.

\section{Model}
{Let us denote a site on a ring (a one-dimensional lattice with the periodic boundary condition) by an index $i\in \{1,\ldots,L\}$, and a species on site $i$ by $\sigma_i\in \mathcal{S} \equiv \{1,\ldots , N \}$ with $N\ge3$}.
Each species has one prey and one predator through a cyclic competition as shown in {Fig.~\ref{ModelFig}}.  
Explicitly, we consider a continuous time Markov process,
where the species at site $i$ with $\sigma_i=\alpha$
invades the nearest neighbor sites $j\in\{i-1,i+1\}$ 
with rate $1/2$ only if $\sigma_j = \alpha_+$
where $\alpha_\pm \equiv \alpha \pm 1~{\rm modulo}~N$, making $\sigma_j=\alpha$ 
after invasion. In addition to this invading process, 
we introduce the exchange of two individual species
on a uniformly chosen random pair of two sites $i,j$ at rate $v/2$ {per one site}
as global mixing \cite{akhmetzhanov2013effects}: Namely, $(\sigma_i,\sigma_j)=(\alpha,\beta)$ with $\alpha, \beta\in \mathcal{S}$ 
is updated as $(\sigma_i,\sigma_j)=(\beta,\alpha)$ after mixing.
Note that this mixing process does not change the population {density} of each species in the system \cite{commentsCopy}.
{The initial probability of finding each species $\alpha$ on site $i$
  is assumed to be determined by $Q(\alpha)$, independent of site $i$, namely with $\sum_{\alpha\in\mathcal{S}}Q(\alpha)=1$.}

We define the $n$-point probability $P_{\{\alpha_k\}_{k=1}^{n}}(t)$ 
as the probability of a randomly chosen segment of $n$ sites 
containing the sequence of species $\{\alpha_k\}_{k=1}^n$, in this order from the left to the right,
with $\alpha_k\in\mathcal{S}$ for any integer $k$, $1\le k\le n$ at time $t$.
{The time evolution of the one-point probability can be determined
  by taking into account all possible changes of species in invasion processes at pairs of adjacent sites.
Then, the one-point probability is described by the following two-point probabilities}
\begin{equation} \label{eq:one}
\dot P_{\alpha} = \frac{1}{2}(P_{\alpha,\alpha_+}+P_{\alpha_+,\alpha}-P_{\alpha_-,\alpha}-P_{\alpha,\alpha_-}).
\end{equation}
In the limit of $L\to\infty$ and $v\to\infty$, one can easily find that this becomes the LV equation (\ref{LV}). 
Thus, in general, the evolution equation of $(n-1)$-point probability
is exactly described by only the $n$-point probability \cite{frachebourg1998fixation}.
Note that Eq. (\ref{eq:one}) does not explicitly depend on the mixing rate $v$ due to the conservation of the populations under mixing,
while explicit contributions from mixing arise in $\dot P_{\{\alpha_k\}_{k=1}^n}$ for any $n\ge2$. 
\section{Results}
\subsection{Perturbative approach to deterministic extinction}
We {employ} a perturbation approach from the infinite mixing rate
where the limit of $L\to\infty$ is taken first.
In order to obtain a closed set of evolution equations,
it is key to realize the following relation for any integer $n\ge2$:
\begin{equation} \label{app0}
P_{\{\alpha_k\}_{k=1}^{n}} = \frac{P_{\{\alpha_k\}_{k=1}^{n-1}} P_{\{\alpha_k\}_{k=2}^{n}}}{P_{\{\alpha_k\}_{k=2}^{n-1}}}+\mathcal{O}(v^{-1-2(n-2)}),
\end{equation}
{where $P_{\{\alpha_k\}_{k=2}^{1}}\equiv 1$ and $\mathcal{O}(x^a)$
  has a finite real number $C$ independent of $x$ such that $\mathcal{O}(x^a)/x^a \le C$ when $x\ll 1$.
The term $O(v^{-1})$ comes from the fact that in order for $\alpha_1$ to influence $\alpha_N$,
initially species $\alpha_1$ at the most left site has to invade $\alpha_2$ before mixing to make
predator $\alpha_1$ away. The probability that such an event occurs is $\mathcal{O}(1/v)$.
Then, the rest of the term $\mathcal{O}(v^{-2(n-2)})$ comes from the fact that 
{$\alpha_1$ must succeed to invade $\alpha_n$ through the segment of $(n-2)$-sites to influence $\alpha_n$ at the most right site by invading all species in between
and $\alpha_1$ at the most left site must not be affected by the mixing.}
The probability that such an event occur is $\mathcal{O}(v^{-2})$ at each invasion,
leading to $\mathcal{O}(v^{-2(n-2)})$ totally.
Note that Eq. (\ref{app0}) with $n=3$ ignoring $\mathcal{O}(v^{-1-2(n-2)})$ was
  used as (Kirkwood) pair approximation in order to analyze the case of $v=0$ \cite{tainaka1994intrinsic,tainaka1994vortices,frachebourg1998fixation}.}

Using Eq. (\ref{app0}) with $n=3$ and ignoring the term of $\mathcal{O}(v^{-3})$,
we obtain the following equation for any $\alpha,\beta\in\mathcal{S}$:
\begin{eqnarray} \label{eq:two}
&\dot P_{\alpha,\beta} =\frac{1}{2} \bigg( - \delta_{\beta,\alpha_+} \frac{P_{\alpha,\beta}}{P_{\beta}} \sum_{\gamma \in \mathcal{S} \setminus \{\beta \}} P_{\beta,\gamma} - \frac{P_{\alpha,\beta}P_{\beta,\beta_-}}{P_{\beta}} \nonumber\\
&-\frac{P_{\alpha_-,\alpha}P_{\alpha,\beta}}{P_{\alpha}}- \delta_{\alpha,\beta_+}\frac{P_{\alpha,\beta}}{P_{\alpha}} \sum_{\gamma \in \mathcal{S} \setminus \{\alpha\}} P_{\gamma,\alpha} \nonumber \\
 &+ \frac{P_{\alpha,\alpha_+}P_{\alpha_+,\beta}}{P_{\alpha_+}}  (1-\delta_{\beta,\alpha_+}) + \frac{P_{\alpha,\beta_+}P_{\beta_+,\beta}}{P_{\beta_+}}  (1-\delta_{\alpha,\beta_+})\nonumber \\
  &+ \delta_{\alpha,\beta}(P_{\alpha,\alpha_+} + P_{\beta_+,\beta}) \bigg) \nonumber\\
  &+ 2 v \left( P_{\alpha} P_{\beta}  -  P_{\alpha,\beta}   \right),
\end{eqnarray} where {$\delta_{a,b}$ is the Kronecker delta function
  and $\mathcal{S}\setminus \{\alpha\}$ denotes the set {obtained by} excluding $\alpha$ from set $\mathcal{S}$.}
Thus, Eq. (\ref{eq:one}) and Eq. (\ref{eq:two}) give a closed set of equations,
allowing us to calculate exactly $P_\alpha$ and $P_{\alpha,\beta}$
within the errors of $\mathcal{O}(v^{-3})$. {It should be noted that, precisely speaking,
  we should add that the reliability of the calculation holds up to only the time regime $t\ll \mathcal{O}(v^3)$ for sufficiently large values of $v$
  because of the possible accumulation of error terms.}

We now move on to introduce an indicator:
\begin{eqnarray}
\eta(t)\equiv \min_{\alpha\in{\mathcal S}}P_\alpha(t), 
\end{eqnarray} in order to discuss the extinction of, at least, one species, which is defined as $\eta\to 0$ as $t\to\infty$.
Indeed, instead of directly focusing on $\eta$,
it turns out that it is convenient to consider a quantity {$E\in[0,1]$} representing the intrinsic cyclic symmetry:
\begin{equation}
E(t)\equiv {N^N}\prod_{\alpha \in \mathcal{S}}P_{\alpha}(t),
\end{equation} because $E=0$ corresponds to $\eta=0$ and thus one may conclude that extinction occurs {if} $E\to 0$ as $t\to\infty$.
One can easily find that $E$ is a conserved quantity in the limit of $v\to\infty$ by {directly} analyzing the LV equation (\ref{LV}) \cite{hofbauer1998theory, Zia2011,Frey2013}.

Let us estimate $\dot E$ within $\mathcal{O}(v^{-2})$
by {using Eq. (\ref{app0}) with $n=2$ for} $P_{\alpha,\beta}$ in the following way:
\begin{equation} \label{app1}
P_{\alpha,\beta} = P_{\alpha}P_{\beta} + \Delta_{\alpha,\beta}v^{-1} + \mathcal{O}(v^{-2}),
\end{equation} where the order of $\Delta_{\alpha,\beta}$ is $\mathcal{O}(v^0)$ due to Eq. (\ref{app0}).
By substituting Eq. (\ref{app1}) into Eq. (\ref{eq:two}), one obtains the explicit expression of $\Delta_{\alpha,\alpha_+}$ as
\begin{eqnarray}
 { \Delta_{\alpha,\alpha_+}=\frac{1}{4} P_\alpha P_{\alpha_+}(P_{\alpha_-} + P_\alpha - P_{\alpha_+} - P_{\alpha_{++}}-1),}\label{Delta}
\end{eqnarray} {and relation $\Delta_{\alpha,\alpha_+} = \Delta_{\alpha_+,\alpha}$.}
Then, one may finally obtain
\begin{eqnarray}
\dot E &=& 
- \frac{1}{4} E \sum_{ \alpha \in \mathcal{S}}\big( P_{\alpha_-} - P_{\alpha_+} \big) ^2 v^{-1} + \mathcal{O}(v^{-2}).
\label{Fexpand}
\end{eqnarray}

{Assuming} that this perturbation series has a {nonzero-radius of convergence} in terms of $v^{-1}$,
let us elaborate on what Eq. (\ref{Fexpand}) means in terms of extinction. 
Indeed, $\dot E$ is always non-positive and {can be zero only}
if the state is extinct with $E=0$ or balanced with $P_{\alpha_-}=P_{\alpha_+}$
for any $\alpha\in\mathcal{S}$. Further, $E$ takes the maximum value {$1$}
{only} when the state is symmetric with $P_\alpha=1/N$ for any $\alpha\in \mathcal{S}$.
{Note that in the case of odd species number, this symmetric condition is the only case to satisfy 
  the balanced condition $P_{\alpha_+}=P_{\alpha_-}$.
  Therefore, we may conclude $E\to 0$ as $t\to\infty$
unless the initial condition satisfies $Q(\alpha_-)=Q(\alpha_{+})$ for any $\alpha\in\mathcal{S}$.}

{In the case of even species number,
there are asymmetric balanced states with $P_{\alpha_-}=P_{\alpha_+}>0$ and $P_{\alpha}\neq P_{\alpha_+}$
for any $\alpha\in\mathcal{S}$ where $E\notin \{1,0\}$.}
{Therefore, such nontrivial asymmetric balanced points might affect the trajectories of states.
In order to clarify this point, we consider Lotka-Volterra equation (\ref{eq:one})
combined with Eq. (\ref{app1}) and Eq. (\ref{Delta}),
which leads to an evolution equation (perturbative-LV equation) closed by only one-point probabilities.}
{Indeed, asymmetric balanced points in the perturbative-LV equation are fixed points,
  which can be simply characterized as a set of $(q_1,q_2)$ where we put $P_\alpha=q_1\le P_{\alpha_+}=q_2$
  for a odd species number $\alpha$ without loss of generality.
Therefore, the Jacobian matrix at the asymmetric balanced points has 
a certain circulant property. Namely, the Jacobian matrix can be constructed
by only three different $2\times2$ matrices as entries
and this Jacobian matrix is circulant in terms of those entries.}
{Note that the analytical expressions of eigenvalues for any circulant matrix
  are well-known independent of the matrix sizes \cite{circulant}.}

{Therefore, using a standard method for usual circulant matrices with taking into account such a quasi-circulant property,
  one may obtain analytical expression of those $N$ eigenvalues $\{\lambda_{k}^\pm\}_{k=1}^{N/2}$ for any even species number $N$ as follows:}
\begin{eqnarray}
  \lambda_k^{\pm} &=& 2\epsilon q_1q_2(1-\Re(\zeta^k))\nonumber \\
  &\pm& \boldsymbol{i}\sqrt{2q_1q_2(1-\epsilon)^2(1-\Re(\zeta^k))} +\mathcal{O}(\epsilon^{2}),\label{aeigen}
\end{eqnarray} {where $\boldsymbol{i}$ is the imaginary unit,
  $\epsilon\equiv v^{-1}/4$, $\zeta=\exp(\boldsymbol{i}4\pi /N)$,
  $k$ is integer such that $1\le k\le N/2$, and $\Re(A)$ is the real part of a complex number $A$.
   An important point is that $\Re(\lambda_k^\pm)>0$ for $k\neq N/2$
   because $(1-\Re(\zeta^k))\ge 0$ and the equality holds only when $k=N/2$.
   This zero eigenvalue is a trivial consequence of $\sum_{\alpha\in\mathcal{S}}P_\alpha=1$.
  This means that the asymmetric balanced fixed points with $P_\alpha=q_1\gg\epsilon$ are unstable.
  Note that in the case of $\epsilon=0$ and $q_1=q_2$, one may directly
  derive $\lambda_k^\pm=\pm\boldsymbol{i}(2/N)\sin(2\pi k/N)$ from Eq. (\ref{aeigen}),
  which is consistent to the eigenvalue of the LV equation (\ref{LV})
  at the fixed point with the symmetric condition for general species number $N$ \cite{hofbauer1998theory}.}

{Summarizing the above results, within the perturbation calculation in terms of $v^{-1}$,
  we  conclude that for any species numbers $N\ge 3$,
  $E\to 0$ as $t\to\infty$ unless the initial condition satisfies $Q(\alpha)=Q(\alpha_{++})$  and $Q(\alpha_-)=Q(\alpha_{+})$ for any $\alpha\in\mathcal{S}$.
This leads to $\eta\to 0$ as $t\to \infty$ under the same condition.  
  Note that under the discussion until here, the fixed point with $P_\alpha=1$ and $P_\beta=0$ for any $\beta\neq\alpha$ is one candidate of the state with $E=0$.
  However, the eigenvalues at this fixed point in the limit of $\epsilon=0$ take only three values from $\{1,-1,0\}$
  for general species number $N$, meaning that this fixed point with $E=0$ is still unstable for sufficient small values of
  $\epsilon$. Thus, {next, we examine the different types of trajectories that appear as} $E$ approaches $0$.}

\subsection{Numerical analysis of extinction trajectories and bifurcation}

\begin{figure}
  \includegraphics[width=0.5\textwidth]{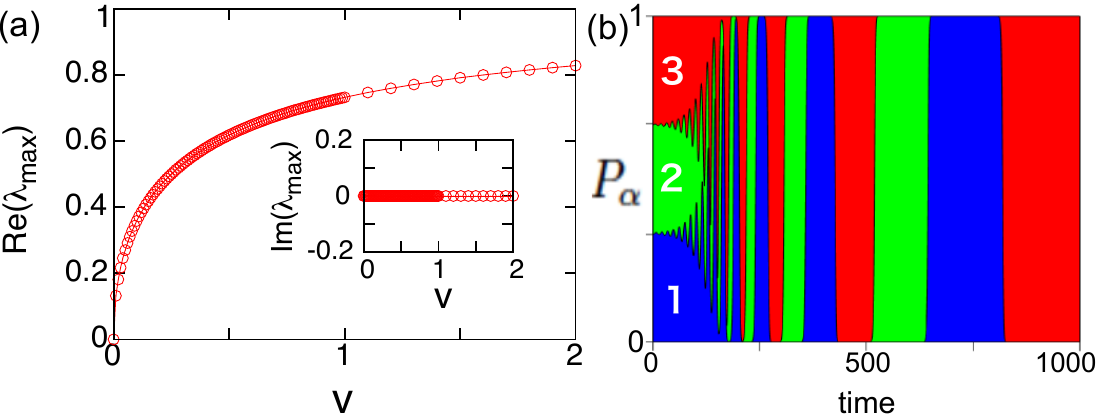}
  \caption{
    \label{N3plots} (color online)
    (a) The maximum real part of the eigenvalues $\lambda$ at the {extinction} fixed points for $N=3$.
    The imaginary part is shown in the inset.
    (b) $P_\alpha(t)$ computed by Eqs.~(\ref{eq:one}) and (\ref{eq:two}) for $v=1$
    from the symmetric initial condition with a perturbation.
  }
\end{figure}

We next discuss through which trajectory the system approaches the extinction $\eta\to 0$,
in particular, by employing the linear stability analysis of Eq.~(\ref{eq:one}) and (\ref{eq:two})
at both fixed points $P_\alpha P_{\alpha_+}=0$ for any $\alpha\in\mathcal{S}$ and $P_\alpha=1/N$ for any $\alpha\in\mathcal{S}$,
with $\dot{P}_\alpha=0, \dot{P}_{\alpha,\beta}=0$ for any $\alpha,\beta\in\mathcal{S}$.
{For later convenience, we call the former fixed point where at least one species has zero population as an {\it extinction fixed point}, and the latter fixed point as a {\it symmetric fixed point}.
Note that one can easily reduce the number of variables from $N+N^2$ to $N^2-N$ for any species number $N$ by using conservation laws such as $\sum_{\alpha\in\mathcal{S}}P_\alpha=1$.
Therefore, we actually deal with $(N^2-N)\times(N^2-N)$ Jacobian matrix.}
{In contrast to the perturbative Lotka-Volterra equation derived in the previous section,
  the evolution equations closed by one-point and two point probabilities have more information about correct trajectories,
  especially for small values of $v$.}
However, such a linear stability analysis for {rather small values of} $v$ should be referred to as an approximation,
nevertheless, it would give us insightful information as found below.

We computed
  $P_{\alpha,\beta}$ at the symmetric fixed points by solving 
  Eqs.~(\ref{eq:one}) and (\ref{eq:two}) numerically from the initial condition where
  $P_\alpha=1/N, P_{\alpha,\beta}=\delta_{\alpha,\beta}/N$. By using this initial
  condition with no invasions, one can avoid coarsening effects causing slow
  relaxations. At the {extinction} fixed points, one can obtain the exact relation
  $P_{\alpha,\beta}=P_{\alpha}P_\beta$.
Hereafter, we call the eigenvalue with the largest real part among all the eigenvalues as the dominant eigenvalue
and the corresponding eigenvector as the dominant eigenvector.
If the real part of the dominant eigenvalue is positive, the dominant eigenvector is {also referred to as} the most unstable eigenvector.
Let us discuss the cases of $N=3$, $N=4$, and also $5\le N\le8$ in this order below.

In the case of 3-species competition ($N=3$),
there are three extinction fixed points {where two species have zero populations, namely}
$P_\alpha=1$, $P_{\alpha_+}=P_{\alpha_-}=0$ for any $\alpha\in\mathcal{S}$.
The linear stability analysis indicates that the dominant eigenvalue is positive at any finite $v$ as shown in Fig.~\ref{N3plots}a.
Thus, the {extinction} fixed points turn out to be saddles.
Therefore, one consistent way to realize the trajectory to extinction satisfying $\eta\to 0$
is a heteroclinic cycle connecting such unstable {extinction} fixed points.
As shown in Fig.~\ref{N3plots}b, by solving numerically Eq. (\ref{eq:one}) and Eq. (\ref{eq:two}) 
from the initial conditions close to the symmetric fixed point,
we indeed observed such heteroclinic cycles.
This kind of trajectories was also observed by looking at the species densities in MC simulations 
 at rather small $v$. This implies that mixing always causes the deterministic extinction for $N=3$ even with small $v$,
 leading to the heteroclinic cycle.

\begin{figure}
  \includegraphics[width=0.5\textwidth]{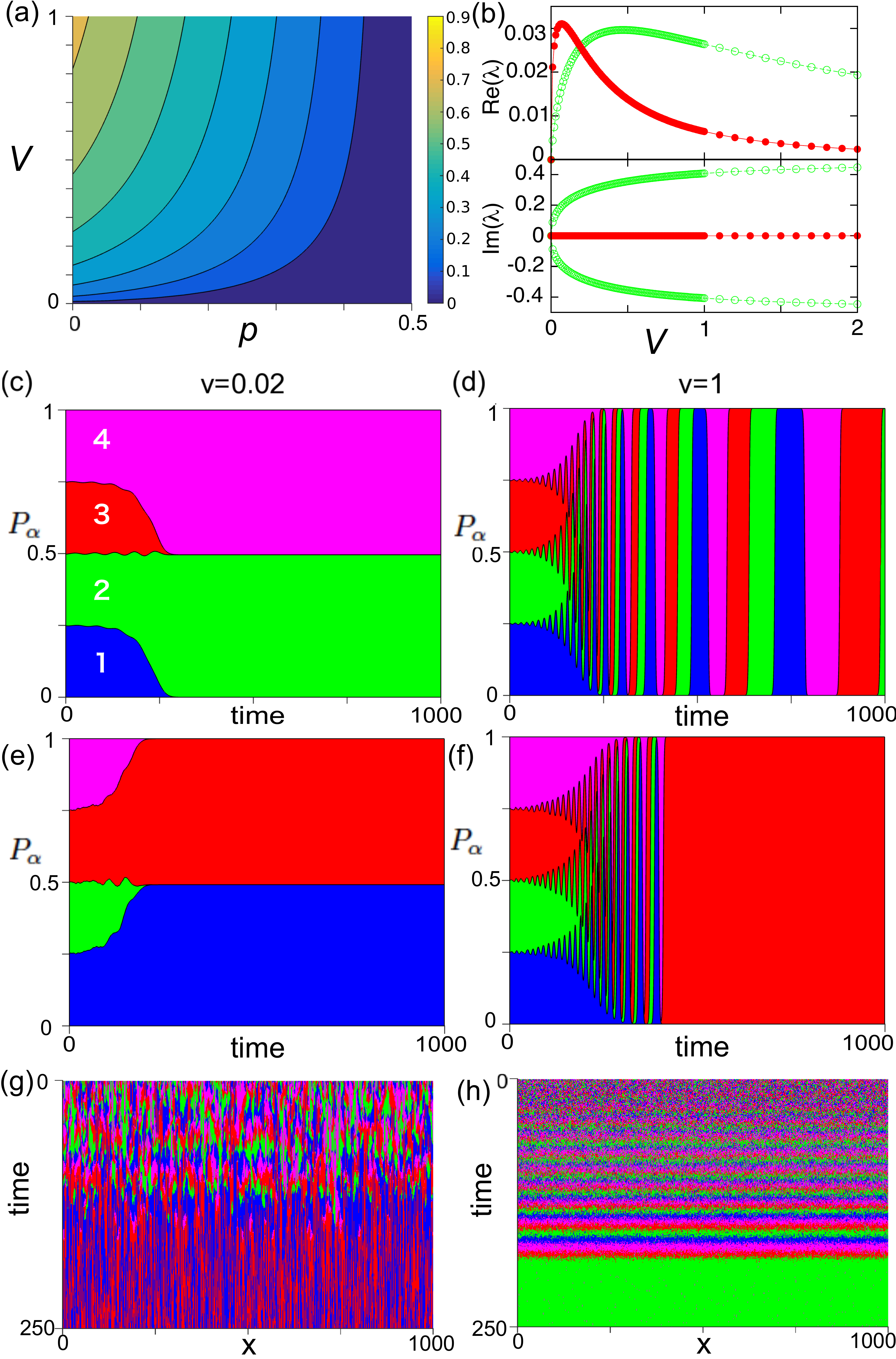}
  \caption{
    \label{N4plots} (color online)
    (a) The maximum real part of eigenvalues at the {extinction} fixed point for $N=4$ as a function of $p$.
    (b) Eigenvalues $\lambda$ at the symmetric fixed point with $N=4$. 
    The top panel shows the two eigenvalues with the largest and the second largest real part,
    and the bottom panel shows their respective imaginary parts.
    (c) $P_\alpha(t)$ computed by Eqs.~(\ref{eq:one}) and (\ref{eq:two}) with $N=4$
    from the symmetric initial condition with a perturbation for $v=0.02$.
    (d) $P_\alpha(t)$ computed by Eqs.~(\ref{eq:one}) and (\ref{eq:two}) with $N=4$ 
    from the symmetric initial condition with a perturbation for $v=1$.
    (e-f) MC simulations: The species densities with $L=9000$ (e,f) and the spatiotemporal plot with $L=1000$ (g,h) for $v=0.02$ (e,g) and $v=1$ (f,h), where $x$ is site number.}
\end{figure}

In the case of 4-species competition,
there are two connected regions parameterized by a real number $p\in [0,1]$,
consisting of {extinction} fixed points {corresponding to} $P_\alpha=p$, $P_{\alpha_{++}}=1-p$, and $P_{\alpha_+}=P_{\alpha_-}=0$
for {$\alpha=1,2$ where survival of two species is allowed}.
As shown in Fig.~\ref{N4plots}a, we found that there is a region in the phase space, close to $p=1/2$,
where only the dominant eigenvalue is zero, meaning that all the rest eigenvalues have negative real parts.
Since the direction along the connected {extinction} fixed points in the phase space is obviously neutrally stable,
corresponding the zero eigenvalue, these {extinction} fixed points around $p=1/2$ compose an attractor.
As $v$ is increased, the {extinction} fixed points for smaller $p\le 1/2$ lose its local stability,
leading that this extinction attractor disappears for any $p$ in the limit of $v\to\infty$. 
{Also, the stability analysis of the symmetric fixed point revealed that
  as $v$ is decreased from a sufficiently large value,
  at $v= v_{\rm c}\approx 0.19$, the most unstable eigenvector at the symmetric fixed point changes
from an oscillatory mode to a non-oscillatory mode (Fig.~\ref{N4plots}b).
Indeed, the numerical solutions of Eq. (\ref{eq:one}) and Eq. (\ref{eq:two})
from the initial conditions close to the symmetric fixed point provide more information about what happen near
$v=v_{\rm c}$ as follows.}
For $v=1>v_{\rm c}$, the trajectories seem to end up with a heteroclinic cycle as shown in Fig.~\ref{N4plots}d,
implying that the basin of the {extinction} attractor around $p=1/2$ is far from
{the trajectories starting from the symmetric fixed point toward the heteroclinic cycle.}
However, for $v=0.02<v_{\rm c}$, trajectories {mostly} fall into the {extinction} attractor
as shown in Fig.~\ref{N4plots}c. Thus, at the bifurcation point, from the viewpoint of the phase space,
{the newly appearing trajectory starting from} the symmetric fixed point seems to {collide with}
a boundary of the basin of the {extinction} attractor{;
thus, this bifurcation is a global bifurcation \cite{dyn}}.
We also observed these two different types of trajectories by MC simulation as shown in Fig.~\ref{N4plots}(e-f).

We performed the linear stability analysis at the symmetric fixed point for also larger $N$. 
In the cases of $N=5,6,7,8$, we {also} found {switching of}
the most unstable eigenvector at certain values of $v$.
In the case of even $N$, as observed in $4$ species,
there is an eigenvector with zero imaginary part for any finite $v$,
that becomes the most unstable eigenvector through a bifurcation as $v$ is decreased.
In the case of odd $N$, the dominant eigenvector has non-zero imaginary part for any finite $v$.
{Importantly, in both cases of even and odd species number $N\ge 4$,
  we numerically observed that a global bifurcation occurs
  as $v$ is decreased from a sufficiently large value. Below the bifurcation,
  the system converges to an extinction fixed point where several species survive
  as observed also in 4-species case.
  Specifically, in typical numerical experiments
  from the symmetric initial condition with a perturbation,
  just $\lfloor N/2 \rfloor$ species have nonzero populations
  at the reached extinction fixed points in the case of species number $N$.}
Note that for $N=3$, we did not find any {switching of} the most unstable eigenvector.
Finally, for general $N$ as far as we have studied, the real part of the dominant eigenvalue
is non-negative and approaches zero only in the limit of both $v\to0$ and $v\to\infty$.
Thus, from the viewpoint of this instability of the symmetric fixed point,
the {two} limits of $v\to0$ and $v\to\infty$ are rather singular.

\section{Concluding remarks}
In this paper, we have studied a cyclically competing species model on a ring with a global mixing.
With a perturbative analysis combined with numerical methods,
we have clarified that the deterministic extinction occurs for {any species number $N\ge 3$ at rather general mixing rate} $v$
and its singular appearance from the {two} limits of $v\to0$ and $v\to\infty$
is a key to bridge the gap between two descriptions of the LV equation and lattice models.

{It would be natural to ask how the dimension of a lattice affects the dynamics of the system with invasion rate $1/2d$.
  Indeed, in general, the deterministic extinction at sufficiently large mixing rates
  is expected to remain in higher dimensions because for such large mixing rates,
  the system would behave as if invasion rate was unchanged compared to that in one dimension.
  Nevertheless, since it is also expected that some dynamical behaviors different from those in one dimension occur for rather small mixing rates in higher dimension,
  it is intriguing to perform the detailed numerical and analytical studies in higher dimensions.
  The current analysis for one dimension could be generalized to that for higher dimensions.
  As a key point for such further studies, in order to compute the dynamics of the two-point probabilities
  as an exact perturbation in terms of $v^{-1}$, it seems to be necessary to consider $2d+1$ point probabilities which comes
  from one site and its $2d$ neighbor sites, instead of three point probabilities.}

The global mixing studied in this {paper} is not only of theoretical interest. 
For example, a bacterial system consisting of toxin-producing, sensitive, 
and immune bacteria strains can form a cyclic competition
and the mixing process in a liquid culture could be effectively global because of a vigorous mixing \cite{kerr2002,czaran2002chemical}.
The present results in this paper imply that the effects of deterministic extinction could be rather widely observed in experiments
because of the difficulties to realize precisely a well-mixed condition in reality.
In experiments, we expect that the effects from deterministic extinction are observed
when each species density is sufficiently far from zero,
and eventually one species goes extinct because of intrinsic stochasticity in the {finite-size} system
when the minimum species density gets close to zero compared to the strength of the stochasticity.

Lastly, let us mention the universality of the deterministic extinction
{in the presence of a mixing process} in terms of different {spatial} descriptions.
Indeed, {the deterministic extinction by mixing similar to the behavior studied in this paper has been observed 
in a lattice model of 3-species cyclic competition under a flow at fast mixing \cite{karolyi2005rock},}
and also in the framework of the continuous limit described by a partial differential equation of populations
under a turbulent convective flow at fast mixing \cite{grovselj2015turbulence}.
Such observations of deterministic extinction in different {spatial} descriptions
imply the universality of deterministic extinction in cyclic systems with fast mixing, to some extent. 
Thus, we hope that the obtained results in this paper could provide a fundamental viewpoint
to develop our understanding of mixing-induced deterministic extinction in more general ecosystems.

\section{acknowledgments}
The authors thank E. Frey for useful discussions on the topic related to Ref. \cite{grovselj2015turbulence}
and {also Y. Ishitsuka for telling us about the eigenvalues of a matrix with a quasi-circulant property}.
This work is supported by the Danish National Research Foundation. 


\begin{thebibliography}{99}
\bibitem{commentsCopy}%
  As a mixing process, one can consider a copy-paste process where a species at a randomly chosen site
  invades a species at another randomly chosen site with rate $v$.
  In this case, the main results in this paper are unchanged {although each species density is often not conserved by each mixing process}.
\end{thebibliography}
%

\end{document}